\documentclass[a4paper,10pt]{scrartcl}
\usepackage[T1]{fontenc}
\usepackage[english]{babel}
\usepackage{amsmath}
\usepackage{amssymb}
\usepackage{amsfonts}
\usepackage{stmaryrd}
\usepackage[utf8]{inputenc}
\usepackage{natbib}

\begin{document}

\author{T. Hartung}
\title{On radial motion of dense regions and objects in protoplanitary-like disks}
\maketitle

\begin{abstract}
Astronomers recently discovered a non-trivial radial motion of stars within
our galaxy and were able to reproduce the results using n-body numerics. I
modeled a protoplanitary-like disk within a newtonian $\infty$-body problem and
considered the stability of local trajectories giving a qualitative explanation
to the observed radial motion. As a prospect there will be a short estimate on
the impact on spiral arms.
\end{abstract}

\section*{Introduction}
\paragraph{}
Observational and numerical data (\citet{ros}) stress the dynamical consideration of spiral arms as for stars have been observed to migrate radially through the spiral arms leaving significant distances between their observed positions and birthregions. It is well known that spiral arms are trajectories of local energetic minima with a very small tangental increase of potential energy thus making it possible to migrate long distances with low effort. This is why local barycenters migrate along these trajectories. The focus of this paper will now be the dynamics of objects in regions of these local barycenters.

\section*{The model}
\paragraph{}
The model contains of two parts: the local gravitational impact and the non-local gravitational impacts. To consider the local impacts I assume the disk to be hot enough in order to take a region with radius $R$ of a dense region or object (a star or planet to be build) so that the region will be nearly homogenous (density $\varrho$, mass $m_2=\frac{4\pi\varrho R^3}{3}$) in all three dimensions. The potential $U$ then integrates
\begin{equation}
 \Delta U=4\pi\gamma\varrho
\end{equation}
Obviously this leads to
\begin{equation}
 U_i=\frac{2\pi\gamma\varrho}{3}r^2-\frac{\mathcal{A}}{r}+\mathcal{B}
\end{equation}
Demanding $|U_i(r=0)|<\infty$ leads to $\mathcal{A}=0$. Further there shall be 
\begin{equation}
 U_i(r=R)=-\frac{\gamma m_2}{R}\ \Rightarrow\ \mathcal{B}=-2\pi\gamma\varrho R^2=-\frac{3\gamma m_2}{2R}
\end{equation}
\begin{equation}
 \Rightarrow\ U_i=\frac{2\pi\gamma\varrho}{3}r^2-\frac{3\gamma m_2}{2R}=\frac{\gamma m_2}{2R}\left(\frac{r^2}{R^2}-3\right)
\end{equation}
Thus the force will be 
\begin{equation}
 \Rightarrow\ F_i=-\frac{\gamma m_2}{R^3}r
\end{equation}
Taking other forces into account the force will slightly change, hence will be perturbated to
\begin{equation}
 F_i=-\tilde\alpha_2r\qquad,\qquad \tilde\alpha_2>0
\end{equation}
\paragraph{}
Integrating all the non-local impacts you will get a barycentral force 
\begin{equation}
 F_e=-\frac{\tilde\alpha_1}{r'^2}\qquad,\qquad\tilde\alpha_1>0
\end{equation}
With $\alpha_i:=\frac{-\tilde\alpha_i}{m}$, $m$ being the mass of the dense object, Newton's second law becomes
\begin{equation}
\left(\begin{array}{c}
\ddot x\\ \ddot y
\end{array}
\right)=\left(\begin{array}{c}
\frac{\alpha_1x}{\left(x^2+y^2\right)^{\frac{3}{2}}}+\alpha_2\left(x-\chi\right)\\
\frac{\alpha_1y}{\left(x^2+y^2\right)^{\frac{3}{2}}}+\alpha_2\left(y-\psi\right)
\end{array}
\right)
\end{equation}
Whereas $(\chi;\psi)^T$ is the current position of the local barycenter with respect to the barycenter of the non-local masses and $(x;y)^T$ being the position of the dense object with respect to the barycenter of the non-local masses being in the center.
\paragraph{}
I do not consider any motion normal to the plane of the disk, since it is obvious that it will be just some oscillation. The model also only applies to the middle regions of the disk, since there are non-trivial $z$-components to the force inside the bulge and in the outer regions the considered region will not be as homogenous anymore.
\section*{Stability of trajectories}
\paragraph{}
Now the trajectories can be characterized by
\begin{equation}
 f(X):=\left(\begin{array}{c}
 \dot x\\
\dot v_x\\
\dot y\\
\dot v_y
\end{array}
\right)= \left(\begin{array}{c}
 v_x\\
\frac{\alpha_1x}{\left(x^2+y^2\right)^{\frac{3}{2}}}+\alpha_2\left(x-\chi\right)\\
v_y\\
\frac{\alpha_1y}{\left(x^2+y^2\right)^{\frac{3}{2}}}+\alpha_2\left(y-\psi\right)
\end{array}
\right)
\end{equation}
with $X:=(x,v_x,y,v_y)^T$ and $x^2+y^2\gg R^2$.
Obviously is $f$ in $X$ lipschitz continuous and thus complies with the standard prerequisites. A possible Lyapunovfunction will have to integrate
\begin{equation}
\begin{array}{ll}\label{dV}
 0\ge\dot V:=\langle \nabla V|f\rangle=&v_x\partial_xV+\left(\frac{\alpha_1x}{\left(x^2+y^2\right)^{\frac{3}{2}}}+\alpha_2\left(x-\chi\right)\right)\partial_{v_x}V\\
&+v_y\partial_yV+\left(\frac{\alpha_1y}{\left(x^2+y^2\right)^{\frac{3}{2}}}+\alpha_2\left(y-\psi\right)\right)\partial_{v_y}V
\end{array}
\end{equation}
Using the identities $x-\chi=:x'$, $v_x-v_\chi=:v_x'$, $y-\psi=:y'$, $v_y-v_\psi=:v_y'$ and $X':=(x',v_x',y',v_y')^T$ 
\begin{equation}
 V(X'):=\begin{cases}\begin{array}{ll}
\exp\left(-x'^2+v_x'^2-y'^2+v_y'^2\right)-1&,\ x'\ge0,\ v_x'\ge0,\ y'\ge0,\ v_y'\ge0\\
\exp\left(-x'^2+v_x'^2-y'^2-v_y'^2\right)-1&,\ x'\ge0,\ v_x'\ge0,\ y'\ge0,\ v_y'\le0\\
\exp\left(-x'^2+v_x'^2+y'^2+v_y'^2\right)-1&,\ x'\ge0,\ v_x'\ge0,\ y'\le0,\ v_y'\ge0\\
\exp\left(-x'^2+v_x'^2+y'^2-v_y'^2\right)-1&,\ x'\ge0,\ v_x'\ge0,\ y'\le0,\ v_y'\le0\\

\exp\left(-x'^2-v_x'^2-y'^2+v_y'^2\right)-1&,\ x'\ge0,\ v_x'\le0,\ y'\ge0,\ v_y'\ge0\\
\exp\left(-x'^2-v_x'^2-y'^2-v_y'^2\right)-1&,\ x'\ge0,\ v_x'\le0,\ y'\ge0,\ v_y'\le0\\
\exp\left(-x'^2-v_x'^2+y'^2+v_y'^2\right)-1&,\ x'\ge0,\ v_x'\le0,\ y'\le0,\ v_y'\ge0\\
\exp\left(-x'^2-v_x'^2+y'^2-v_y'^2\right)-1&,\ x'\ge0,\ v_x'\le0,\ y'\le0,\ v_y'\le0\\

\exp\left(x'^2+v_x'^2-y'^2+v_y'^2\right)-1&,\ x'\le0,\ v_x'\ge0,\ y'\ge0,\ v_y'\ge0\\
\exp\left(x'^2+v_x'^2-y'^2-v_y'^2\right)-1&,\ x'\le0,\ v_x'\ge0,\ y'\ge0,\ v_y'\le0\\
\exp\left(x'^2+v_x'^2+y'^2+v_y'^2\right)-1&,\ x'\le0,\ v_x'\ge0,\ y'\le0,\ v_y'\ge0\\
\exp\left(x'^2+v_x'^2+y'^2-v_y'^2\right)-1&,\ x'\le0,\ v_x'\ge0,\ y'\le0,\ v_y'\le0\\

\exp\left(x'^2-v_x'^2-y'^2+v_y'^2\right)-1&,\ x'\le0,\ v_x'\le0,\ y'\ge0,\ v_y'\ge0\\
\exp\left(x'^2-v_x'^2-y'^2-v_y'^2\right)-1&,\ x'\le0,\ v_x'\le0,\ y'\ge0,\ v_y'\le0\\
\exp\left(x'^2-v_x'^2+y'^2+v_y'^2\right)-1&,\ x'\le0,\ v_x'\le0,\ y'\le0,\ v_y'\ge0\\
\exp\left(x'^2-v_x'^2+y'^2-v_y'^2\right)-1&,\ x'\le0,\ v_x'\le0,\ y'\le0,\ v_y'\le0\\
\end{array}\end{cases}
\end{equation}
integrates \ref{dV} in the open region 
\begin{equation}
 \Omega:=\left\{(x,v_x,y,v_y)^T:\ |x-\chi|<\frac{\alpha_1x}{\alpha_2(x^2+y^2)^{\frac{3}{2}}},\ |y-\psi|<\frac{\alpha_1y}{\alpha_2(x^2+y^2)^{\frac{3}{2}}}\right\}
\end{equation}
with $V(\chi,v_\chi,\psi,v_\psi)=0=\dot V(\chi,v_\chi,\psi,v_\psi)$ and coordinates respectively chosen to comply $x,y,v_x,v_y,\chi,\psi>0$, which is perfectly possible since $x,y,\chi,\psi>0$ can be chosen as there is $x^2+y^2\gg R^2\ge (x-\chi)^2+(y-\psi)^2$ and $v_x,v_y>0$ define the surface normal. The implicit function theorem ensures that there will always be a time-interval to locally solve these conditions. Should the Jacobian be singular then simply turn the coordinates a little ($x^2+y^2\gg R^2\ge (x-\chi)^2+(y-\psi)^2$). $\Omega$ herein is the largest possible local region to be observed.
\paragraph{}
Obviously 
\begin{equation}
 \forall\left(\begin{array}{c}
x\\v_x\\y\\v_y
\end{array}\right)\not=\left(\begin{array}{c}
\chi\\v_\chi\\\psi\\v_\psi
\end{array}\right):\ \exp\left(-x'^2-v_x'^2-y'^2-v_y'^2\right)-1<0
\end{equation}
thus in every region of $\left(\chi,v_\chi,\psi,v_\psi\right)^T$ exists a point $Z$ with $V(Z)<0$. Hence the local region is instable and as a consequence any dense object within the region with leave this in a finite period of time.
\section*{On migration}
\paragraph{}
Since it is known that the local barycenters travel along the spiral arms any dense object in that regions will move with them and this way migrate radially through the disk. On the other hand each object will leave that region after a finite time and hence travel no longer with that local barycenter. Now the object is no more in a local energetic minimum and hence will move back entering another radially migrating region just to soon leave this again and enter right another. Multiple objects of the same birthregion will do so indepenently. Thus the observed galaxies should show a good mixture of young and old stars and stars with different chemical compositions since there are stars from basically every birthregion close another. These are exactly the observed properties in the middle parts of galaxies.
\paragraph{}
It might be interesting to estimate the density of stars in spiral arms for long term effects. In order to migrate outwards for an object it needs energy which has to be taken from other objects for the sum of energies to stay conserved. Since it seems more likely to transfer low amounts of energy I will approximate the energy-transfer-distribution as gaussian with the expected transfer amount $\mu_E=0$. Let then $T$ be the average time between entering two consecutive regions and let $\Lambda$ be the average radial movement, then the energy-transfer-distribution $\varphi_E$ with standard deviation $\sigma_E$ becomes
\begin{equation}
\varphi_E(\Delta E)=\frac{1}{\sigma_E\sqrt{2\pi}}\exp\left(-\frac{\Delta E^2}{2\sigma_E^2}\right)
\end{equation}
In order to estimate $\sigma_E$ I considered the energy $E_g=\frac{\alpha_1m_3}{2r}$ of a mass $m_3$ at radius $r$. As the radius changes by $\Lambda$ the transfered energy is given by
\begin{equation}\label{mu_E}
 \mu_{\Delta E}=\frac{1}{2}\left(\frac{|\alpha_1|m_3}{2\left(r+\Lambda\right)}+\frac{|\alpha_1|m_3}{2\left(r-\Lambda\right)}\right)=\frac{|\alpha_1|m_3}{4}\left(\frac{2r}{r^2-\Lambda^2}\right)
\end{equation}
Now I average this over the entire applicable region with inner radius $R_I$ and outter radius $R_O$
\begin{equation}
 \overline{\mu}_{\Delta E}=\frac{1}{R_O-R_I}\int_{R_I}^{R_O}\mu_{\Delta E}dr=\frac{|\alpha_1|m_3}{4(R_O-R_I)}\ln\left(\frac{R_O^2-\Lambda^2}{R_I^2-\Lambda^2}\right)
\end{equation}
Now it is possible to approximate $\sigma_E$ by
\begin{equation}
 \frac{1}{2}=2\int_0^{\overline{\mu}_{\Delta E}}\varphi_Ed\Delta E
\end{equation}
The energy-density thus will be homogenous. Considering our own galaxy with $R_I=8000ly\ll r\ll 50000ly=R_O$ one finds $E(8000ly)\approx 6\alpha_110^{-21}\frac{Js^2}{m^3}m_3$ as well as $E(50000ly)\approx \alpha_110^{-21}\frac{Js^2}{m^3}m_3$ and thus $\Delta E=5\alpha_110^{-21}\frac{Js^2}{m^3}m_3$ on a distance of $42000ly$ which is nearly homogenous.
\section*{Conclusion}
\paragraph{}
From (\ref{mu_E}) it is easy to see that the energy-transfer enables much larger distances of migration the farther away an object is from the center of the disk. Hence one will more likely find a higher mixture of objects in the outter regions. On the other hand this gives a mechanism preventing too many masses to fall into the interior regions and thus from possibly falling into black holes. Further keeps the disk from accreting too fast making it possible for planets and stars to form.
\paragraph{}
The local instability hence has an impact on the largescale stability of a protoplanitary disks and spiral arms. It does also explain the observed mixture of stars.

\end{document}